\documentclass[journal,final,twocolumn,letterpaper,10pt]{IEEEtran}

\makeatletter

\let\proof\@undefined
\let\endproof\@undefined

\let\theorem\@undefined
\let\endtheorem\@undefined
\makeatother

\usepackage[active]{srcltx} %
\usepackage{amsmath,amsthm,amsxtra,amssymb}
\usepackage{times,latexsym,array,verbatim}
\usepackage{epsfig,graphicx,cite,times}
\usepackage{epstopdf}
\newtheorem{thm}{Theorem}[section]
\newtheorem{rem}[thm]{Remark}

\newcommand{\wv}{\mathbf{w}}

\newcommand{\uv}{\mathbf{u}}

\newcommand{\xv}{\mathbf{x}}

\newcommand{\zv}{\mathbf{z}}
\newcommand{\muv}{\mathbf{\mu}}

\newcommand{\yv}{\mathbf{y}}

\newcommand{\z}{\| \zv \|^2}

\begin{document}
\title{A Robust Control Framework for Malware Filtering}

\author{Michael~Bloem,
        Tansu~Alpcan,~\IEEEmembership{Member,~IEEE,}
        and~Tamer~Ba\c{s}ar,~\IEEEmembership{Fellow,~IEEE}%
\thanks{M. Bloem is with the NASA Ames Research Center, Moffett Field, CA 94035, USA (e-mail: michael.bloem@nasa.gov). He was with the University of Illinois at Urbana-Champaign and partly supported by Deutsche Telekom AG Laboratories during this project.}
\thanks{T. Ba\c{s}ar is with the Coordinated Science Laboratory, University of Illinois at Urbana-Champaign, Urbana, IL 61801 USA (e-mail: tbasar@control.csl.uiuc.edu).}
\thanks{T. Alpcan is with Deutsche Telekom Laboratories, D-10587 Berlin, Germany (e-mail: tansu.alpcan@telekom.de).}}

\maketitle

\begin{abstract}
We study and develop a robust control framework for malware filtering and network security. We investigate the malware filtering problem by capturing the tradeoff between increased security on one hand and continued usability of the network on the other. We analyze the problem using a linear control system model with a quadratic cost structure and develop algorithms based on H$^\infty$-optimal control theory. A dynamic feedback filter is derived and shown via numerical analysis to be an improvement over various heuristic approaches to malware filtering. The results are verified and demonstrated with packet level simulations on the Ns-2 network simulator. 
\end{abstract}


\section{Introduction}
\PARstart{A}{ttacks} on computer networks, such as worm or denial of services attacks, are difficult to prevent in part due to the challenge of detecting and stopping them while still allowing legitimate network usage. Recent experience with Internet worm attacks makes this point more clear: within 10 minutes the Slammer worm had infected 90\% of vulnerable computers in 2003 and the Code Red virus infected
hundreds of thousands of hosts in 2001~\cite{moore_slammer, moore_red}.  The \emph{base-rate fallacy} captures the essence of this problem. Even if we have low false-negative and false-positive rates in our detection of malware, there is so much more legitimate network usage than illegitimate usage that we end up with many false alarms~\cite{axelsson}.  The incredible variety in legitimate network traffic makes accurately differentiating it from malicious traffic even more challenging.
A more detailed analysis of the detection of a particular type of worm epidemic in ~\cite{rohloff_basar} shows the challenge of detecting some worm attacks even under idealized conditions.  In this specific case the base-rate fallacy again comes into play, as ``a substantial volume of `background radiation'" is to blame for making the detection of
random constant scanning worms difficult. Intrusion detection systems must be constructed with this dilemma in mind, and thus need to be conservative in their operation.

According to Federal Bureau of Investigation (FBI) statistics, 70\% of security problems originate within an organization, and 20\% of respondents to an FBI survey indicated that intruders had broken
into or attempted to break into their corporate networks in the last 12 months~\cite{cisco}.  Therefore, dynamic firewalls such as the Cisco Internetwork Operating System (IOS) firewall are an important form of internal network security~\cite{cisco}. Our aim is to develop algorithms and policies for such (re)configurable firewalls
in order to filter malware traffic such as worms, viruses, spam, and Trojan horses. 

We use $H^\infty$-optimal control theory to determine how to dynamically change filtering rules or parameters in order to ensure a certain performance level.  
We note that in $H^\infty$-optimal control, by viewing the disturbance as an intelligent maximizing opponent in a dynamic zero-sum game, who plays with knowledge of the minimizer's control action, one evaluates the system under the worst possible conditions. This approach applies naturally to the problem of malware response because the traffic deviation resulting from a malware attack is not merely random noise, but represents the efforts of an intelligent attacker. Therefore, we determine the control action that will minimize costs under these worst circumstances~\cite{basarhinf}.
The resulting conservative controller works well even in light of the base-rate fallacy problem.
To the best of our knowledge, this work represents the first application of robust control theory
to the problem of malware filtering.

\subsection{Related work}

There are several methods of dynamic packet filtering~\cite{tulloch}. Perhaps the most common one is to dynamically change which ports are open or closed.  Stateful inspection of deeper layers of packets allows for even more detailed filtering by creating and maintaining information about the state of a current connection~\cite{cisco}. Another possibility is to dynamically alter the set of Internet Protocol (IP) addresses from which traffic will be accepted~\cite{hazelhurst}. An accurate attack packet discarding scheme based
on statistical processing has been proposed in~\cite{kim}, where each packet is associated with a score
that reflects its legitimacy. Once the score of a packet is computed, this scheme performs score-based selective packet discarding where the dropping threshold is dynamically adjusted based on the score distribution of recent incoming packets and the current level of system overload.

Implicit to the network traffic filtering problem considered in this article is the partitioning of a computer network into various sub-networks for administrative and security purposes.  This approach is common, and a separate firewall is often assigned to each sub-network. Zou et al. \cite{zou} have proposed a ``Firewall Network System" based on this very concept. Cisco recommends their IOS firewalls for defending particular sub-networks or LANs in a corporate network~\cite{cisco}. In ~\cite{chen_quarantine}, quarantining these sub-networks is considered as a strategy to slow
the spread of worm epidemics. We note that although the algorithms developed in this paper can be helpful for configuring dynamic firewalls such as the ones described above, our main objective is to develop mathematical foundations and algorithms for future security systems which will be even more configurable and flexible.  Finally, while we consider the case of filtering packets, these techniques could also be applied to filtering connections.

The remainder of this article is structured as follows: Section~\ref{sec:filtering} discusses the problem of filtering
network traffic with dynamic firewalls separating sub-networks.  We next derive the $H^\infty$-optimal controller and state estimator in Section~\ref{sec:h_inf_controller}. Section~\ref{sec:filtering_numerical} reviews Matlab and Ns-2 simulations of the H$^{\infty}$-optimal controller and demonstrates its performance in comparison with other controllers. 
Concluding remarks and directions for future research are presented in Section~\ref{sec:h_inf_conclusion}.

\section{Network Traffic Filtering Model} \label{sec:filtering}

In this section we present a linear system model for malware traffic 
and study the problem of filtering network traffic to prevent malware
propagation.
Consider a computer network under the control of a single
administrative unit, such as a corporate network.  Assume the
network is divided into sub-networks for administrative and security
purposes~\cite{cisco}.  While we will describe the model within this
context, the corresponding control framework can be applied
to other contexts by redefining the entities in question.

Let $x(t)$ represent the number of malware packets that traverse a link on their way to the
destination sub-network at time $t$ originating from infected sources outside the sub-network.
We model this malware flow to the sub-network using a linear differential equation
with control and disturbance terms:
\begin{equation} \label{e:dynamicsscalar}
 \dot x(t) = a\, x(t) + b\, u(t) + w_a(t),
\end{equation}
where $u(t)$ represents the number of packets that are filtered at a particular
time $(t)$.  Usually, only some proportion of the packets filtered are actually malware related. Thus, the parameter $b$ corresponds to that proportion multiplied by $-1$. 
In other words, $(1-b)$ is the proportion of filtered packets that are not malware related.
On the other hand, $w_a(t)$ represents the number of malware packets added to the link at time
$t$ intentionally by malicious sources or unintentionally by hidden software running on hosts,
both located outside the sub-network considered. Thus $u(t)$ and $w_a(t)$ represent, for this specific sub-network, the packet filtering rate and
malware infiltration rate, respectively. The
$a$ value represents the instantaneous proportion of malware packets
on the link that are actually delivered to the sub-network and is
thus a negative number.  

Expanding the dimensions of  the model in~(\ref{e:dynamicsscalar})
leads to a set of linear differential equations:
\begin{equation} \label{e:dynamics}
 \dot \xv(t) = A\xv(t) + B\uv(t) + D\wv_a(t),
\end{equation}
where $\wv_a$ is defined as the vector of malware packets.
In this case both $A$ and $B$ are obtained simply by multiplying the identity matrix
by $a$ and $b$, respectively. The $D$ matrix imposes a propagation model on the attack and 
quantifies how malware is routed and distributed on this network. 
For the purposes of this paper, it has zeros for its diagonal terms
(intra-sub-network malware traffic does not leave the sub-network),
and each column must sum to 1 to ensure conservation of packets. In this version of the problem, the malware being sent to sub-network $i$ is a function of $w_j$ for $j \neq i$, the malicious traffic generated by other sub-networks.  This assumption on the propagation of malware inherent to the form given to $D$ allows for a centralized filtering solution that considers network-wide conditions.  A decentralized version to this problem is also possible, however.
Overall, this model simplifies actual network dynamics by assuming a linear
system and using a fluid approximation of traffic flow.

Let us denote by $\yv(t)$ our measurement of the number of inbound malicious packets prior to filtering. Note that the separation
between detection ($\yv(t)$) and response ($\uv(t)$) is only at the conceptual level. In the implementation
both may occur on the same device. Inaccuracies in
$\yv(t)$ are inevitable due to the challenging problem of
distinguishing malicious packets from legitimate
ones~\cite{axelsson}. To capture this uncertainty formally, we define
$\yv(t)$ as
\begin{equation} \label{e:app_y_def}
 \yv(t) := C \xv(t) + E \wv_n(t),
\end{equation}
where $\wv_n(t)$ is measurement noise of any form. Later, we derive and apply the worst-case measurement noise $\wv_n(t)$. Additionally, we
define $N:=EE^{T}$ and assume that it is positive definite, meaning
that the measurement noise impacts each dimension of the measured
output.
The $C$ matrix models the assumption that $\yv(t)$ is higher than and
proportional to $\xv(t)$. When
implemented, entries of this constant matrix could be measured from an analysis of
packet filtering and the calculations required for determining the optimal
controller could be rerun periodically.

Note that we do not make any assumption on how $\yv(t)$ is
obtained. It could be the result of some statistical
analysis comparing the expected traffic to the measured traffic or
be based on a set of rules where packets with certain
characteristics are assumed to be malicious.

Similarly, $\wv_a(t)$ represents a worm attack, expressed in terms of
number of the malware packets sent from a sub-network to other
sub-networks at each time instant. 
More precisely, it is the
generated malware traffic flow rate in terms of packets per time
step.
For example, if a worm is very rapidly contacting new hosts
and sending them packets, then $\wv_a(t)$ would be large.
However, we do not assume any form on the attack.
To simplify notation, we assume
that the measurement noise and attack disturbance
are both part of the vector $\wv := \left[ \wv_a^T ~\wv_n^T \right]^T$.

The model at hand contains several simplifications and
assumptions.  As was mentioned earlier, the components of the $B$
matrix are set to be constants, although in reality the value of
these components should change as $\xv$ decreases, as there are less
malicious packets to be filtered, and we are filtering packets we are
less sure about.  This quantity also depends on the amount of
legitimate network traffic on the link: if there is a relatively
large amount of legitimate network traffic then we will incur more
false-positives and thus end up filtering more legitimate traffic.
The $B$ matrix is related to the
false-negative and false-positive ratios, but it is mostly determined by
the ratio of legitimate to illegitimate traffic as described in ~\cite{axelsson}.
The exponential decay in the number of malware packets on the link
(in the absence of control and disturbance) does not exactly capture
network dynamics, but with a high enough rate of exponential decay,
this assumption is quite realistic when capacity constraints are not
significant.  The assumption of a constant value for
the $C$ matrix is also an approximation, as in reality the number of
malware packets prior to filtering will probably not be linearly dependent
upon the number after filtering.  To summarize, this model
simplifies actual network performance by assuming linear dynamics.

Moreover, this model simplifies system dynamics by using a fluid approximation
of traffic flow.  More specifically, this model only approximately captures
the fact that, in an actual implementation, the number of malware
packets measured prior to filtering differs from the one that arrives
at the sub-network in the number of the filtered.
Similarly, in order to simplify the following
calculations, we are approximating a clearly discrete and
event-driven system (a computer network) with a continuous time
system.  This assumption should hold when we consider the rapidity
and frequency of packet arrivals and transmissions along with the
fine-grained time increments of a computer network.

\section{Derivation of Optimal Controller and State Estimator} \label{sec:h_inf_controller}

Our objective now is to design an algorithm or controller for traffic filtering given this imperfect measure of inbound malicious packets. As part of the $H^\infty$-optimal control analysis and design we
introduce first the \emph{controlled output}
\begin{equation}
\label{e:hinf_z}
 \zv(t) := H\xv(t) + G\uv(t) ,
\end{equation}
where we assume that $G^T G$ is positive definite, and that no cost is placed on the product of control
actions and states: $H^{T}G=0$.  $H$ represents a cost on malicious packets arriving at a sub-network.  A few other constraints that must be met for this $H^\infty$-optimal control theory to apply are that $(A,\,B)$ and $(A,\,D)$ be stabilizable, and $(A,\,H)$ and $(A,\,C)$ be detectable, and these conditions readily hold in our case.

If $\xv$ becomes negative, we are filtering legitimate packets from
the link.  In other words, an equal penalty is assumed for
underfiltering and allowing worm-related traffic on a link and also
for overfiltering and preventing legitimate network traffic from
traversing the link. By weighting these two quantities equally, we
are in effect encouraging survivability: overfiltering to prevent
the spread of the worm but at the same time crippling the network is
penalized as much as allowing the worm-related traffic to run
rampant.

The cost on filtering legitimate traffic is actually more
complicated than indicated above.
Recall that $b$ specifies the
proportion of filtered traffic that is malware-related. Thus, $(1-b)$
is the proportion of filtered traffic that is legitimate (assuming
$\xv$ is positive).  If we assign a cost of $f_l$ to filtering
legitimate packets when malware packets are on the link and a cost
of $f_a$ to the filtering action itself, the components $g$ of $G$
can be specified as $g = f_l(1-b) + f_a$.

The cost of this system for the purpose of $H^\infty$ analysis is defined by
\begin{equation} \label{e:hinf_cost}
 L(\xv,\uv,\wv) = \frac{\| \zv \|}{\| \wv \|},
\end{equation}
where $\| \zv \|^{2} := \int_{-\infty}^{\infty}| \zv(t) |^{2} dt$ and a similar definition
applies to $\| \wv \|^{2}$.
This is a cost ratio rather than an actual cost, but we will refer to
it as the cost for simplicity. It captures the proportional changes in $\zv$ due to changes
in $\wv$. More intuitively, it is the ratio of the cost incurred by the system to the corresponding attacker and measurement noise ``effort'.'

There are a few assumptions and simplifications present in this cost structure.  We
assign a cost to the malware packets, not the infected and disabled
hosts or servers themselves, which are the often actually where the
costs of malware occur.  On the other hand, malware traffic itself
can dominate network resources and thus be costly in its own right.
Another assumption is that we assign costs to
traffic incoming to a sub-network even if that sub-network is
already infected, in which case the incoming malicious traffic would
be unimportant.  In spite of these two assumptions, this cost
structure captures most of the important characteristics of malware
packet propagation.

$H^\infty$-optimal control theory not only applies very directly and
appropriately to the problem of worm response, but also guarantees
that a performance factor (the $H^\infty$ norm)  will
be met. 
This norm can be thought of as the worst possible value for the cost
$L$ and is bounded above by 
\begin{equation} \label{e:hinf_optimal_performance}
 \gamma^{*} := \inf_{\uv} \sup_{\wv} L(\uv,\wv),
\end{equation}
which can also be viewed as the optimal performance level in
this $H^\infty$ context.

In order to actually solve for the optimal controller $\muv(\yv)$, the number of packets to filter as a function of the inaccurately measured number of inbound malicious packets, a corresponding differential game is defined between the attackers
and the malware filtering system, which is parameterized by $\gamma$, where $\gamma > \gamma^*$:
\begin{equation} \label{e:hinf_differential_game}
 J_{\gamma}(\uv,\wv) = \| \zv \|^{2} - \gamma^{2} \| \wv\|^{2}.
\end{equation}
The malicious attackers try to maximize this cost function in the worst-case by
varying $\wv$ while the malware filtering algorithm minimizes it via the
controller $\uv$.
A similar application of game theory, where attackers and intrusion detection/prevention
system are modeled as players in a security game, has been investigated in~\cite{alpcancdc04}.

The optimal filtering strategy $\uv=\muv_{\gamma}(\yv)$ can be determined from this
differential game formulation for any $\gamma>\gamma^{*}$.  It is given by ~\cite{basarhinf}

\begin{equation} \label{e:hinf_opt_controller}
 \muv_{\gamma}(\yv) = -(G^{T}G)^{-1}B^{T} \bar Z_{\gamma}\hat \xv,
\end{equation}
where $\bar Z_{\gamma}$ is solved from
\begin{equation} \label{e:GARE_1}
A^{T}Z + ZA - Z(B(G^{T}G)^{-1}B^{T} - \gamma^{-2}D D^{T})Z + H^T H = 0,
\end{equation}
as its unique minimal positive definite solution, and $\hat \xv$ is given by
\setlength\arraycolsep{0.1em}
\begin{eqnarray} \label{e:hinf_worst_state_estimate}
 \dot {\hat \xv} &=& \left[ A - (B(G^{T}G)^{-1}B^{T} - \gamma^{-2}D D^{T})\bar Z_{\gamma}\right]\hat \xv \nonumber \\
 &+& \left[I - \gamma^{-2} \bar \Sigma_{\gamma} \bar Z_{\gamma}\right]^{-1} \bar \Sigma_{\gamma}C^{T}N^{-1} (\yv - C \hat \xv),
\end{eqnarray}
where $\bar \Sigma_{\gamma}$ is the unique minimal positive definite solution of
\begin{equation} \label{e:GARE_2}
A\Sigma + \Sigma A^{T} - \Sigma(C^{T}N^{-1}C - \gamma^{-2}H^{T}H)\Sigma + D D^{T} = 0.
\end{equation}
Here $\hat{\xv}$ is an estimate for $\xv$. This is a linear feedback controller operating on a state estimate.
Further, $\gamma^*$ is the smallest $\gamma$ such that $\rho(\bar{\Sigma_{\gamma}} \bar{Z_{\gamma}}) < \gamma^2$, where $\rho(\Lambda)$ denotes the spectral radius of the matrix $\Lambda$.
The online calculation is
simply a multiplication by the estimate of the system state. Also
note that this controller requires a network-wide knowledge of the
system state estimate and thus this is a centralized control
solution.

There are a few assumptions implicit in this specific controller
formation. The various filters will have to send control packets to
each other, indicating their $\yv$ values.  Moreover, it is assumed
that these filters are able to convert a number of packets to filter per time step
($u(t)$) into a filtering rule that will implement that filtering
rate.  The packets that are most likely to be malicious should be
filtered first.  Exactly how this is done depends on the system
implementation.  For example, a rule-based filter could implement
more rules (block more ports or IP addresses) or  the sensitivity of
an anomaly-based detector could be increased when $u(t)$ increases.

\begin{rem}
 The $H^\infty$-optimal controller derived here
 (\ref{e:hinf_opt_controller}) is a centralized control solution due to
the  $D$ matrix, which imposes a specific malware propagation model.
However, we can apply the same framework to each sub-network separately
by using~(\ref{e:dynamicsscalar}) for each. %
This leads to a decentralized solution consisting of independent scalar $H^\infty$-optimal controllers.
\end{rem}

\section{Simulations} \label{sec:filtering_numerical}

\begin{figure}
\centering
\includegraphics[width=3.25in]{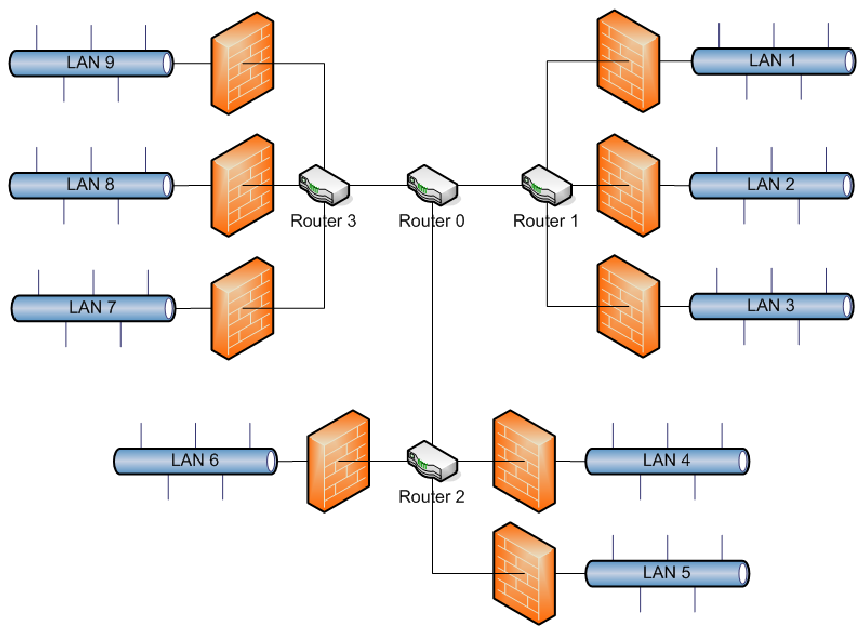}
\caption{Sample computer network to be analyzed.}
\label{fig:system_overview}
\end{figure}

Consider the representative computer network shown in
Fig.~\ref{fig:system_overview}.  In this simple network configuration, each
sub-network or LAN has a dynamic firewall that filters incoming
network traffic. Each firewall communicates its malicious packet measure $\yv$ to all other firewalls, where filtering decisions are made.  No centralized server is overseeing the filtering activity.

\subsection{Simulation setup} \label{sec:sim_setup}

Several attack types are simulated in Matlab on this network topology
in order to compare the
$H^\infty$-optimal controller with other controllers.  As a simplification,
 a sub-network is assumed to be either infected
or not infected.  An infected sub-network sends malware to other
sub-networks.  Sub-networks become infected with some probability
once they have received a certain threshold number of malware
packets.  This probability increases when higher thresholds are met.
Clearly the propagation of these fictitious attacks is much simpler
than that of an actual worm or virus, but it captures the underlying
dynamics of an attack.

Four types of malware attacks are considered: no attack (A1); a
high-traffic, slow spreading attack (A2); a low-traffic,
slow-spreading attack (A3); and a low-traffic, fast-spreading attack
(A4).  In each of these attacks, one subnetwork is initially infected and sends malware to all other sub-networks.

Five response types are applied to each of these four attack types:
no response (R1), the $H^\infty$-optimal controller response (R2), a
threshold-based controller that implements a filter of some fixed
magnitude when a certain amount of malicious packets are detected
(R3), a controller that removes all suspicious packets ($\yv(t)$)
from each link (R4), and an optimal controller that minimizes the
cost $\| \zv \|^2 $ (R5).  For the linear quadratic Gaussian (LQG) optimization problem in (R5),
which is obtained as the limit of the $H^\infty$ problem as
$\gamma \rightarrow \infty$, we use the
expected value of $\int^\infty_{-\infty} \| \zv \|^2 dt$ as the quadratic cost, which we again
denote by $\| \zv \|^2 $ by a slight abuse of notation.

A few details relating to the numerical analysis of these
controllers will now be given.  The $A$ matrix is set to be the
identity matrix multiplied by -1.  Recall that this value quantifies the exponential decay of malicious packets on the link as they arrive at their destination sub-network.  The $b$ quantity is set to 0.5.
This value is consistent with a detection rate (true-positive rate)
of 0.7 and a very low ($10^{-5}$) false-positive rate -- a scenario
considered in ~\cite{axelsson}.  The $D$ matrix is set up such that
sub-networks are more likely to transfer the worm within their group
of three sub-networks.
The $C$ matrix is set to be 2 multiplied by the identity matrix,
which is derived from values observed in the Ns-2 simulations to be
explained in Section~\ref{sec:ns-2}.
It is assumed that $\wv_n$ has
a positive mean, as most malware detection schemes are set up to, if
anything, overestimate the number of malicious packets.  The
standard deviation of $\wv_n$ is relatively low.
Also, the noise is assumed to be
white Gaussian noise, although in reality this noise may well have
some autocorrelation.

Simulations are run with three sets of cost functions ($\| z\|^2$ and $L$)
that differ in their coefficients. The ratio between the cost on inbound
malware packets $\xv$ and the cost on filtering packets $\uv$
(which
involves a cost on filtering legitimate packets and also the
filtering cost itself)
is set at 10:1, 100:1, and 1000:1.

\subsection{Matlab simulations} \label{sec:matlab}

\begin{figure}
\centering
\includegraphics[width=3.25in]{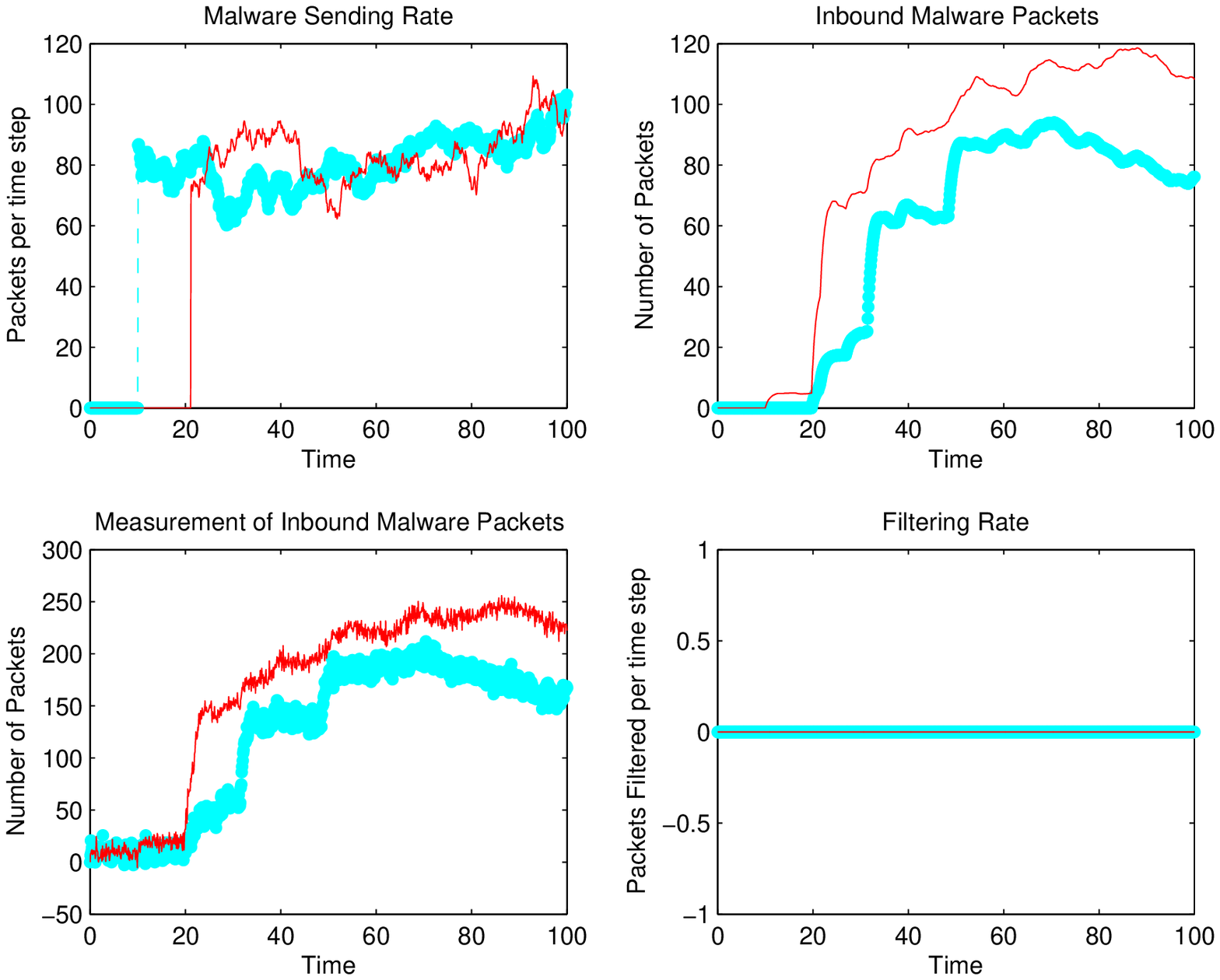}
\caption{\label{fig:no_response} Numerical analysis of slow worm
attack with no response applied on two (out of 9) sub-networks.} \label{resp_none_attack_fast}
\end{figure}

We first conduct a numerical analysis in Matlab. 
The simulations where no response is applied demonstrate that the
assumed malware packet propagation rules  mimic the ``S-shaped" behavior of
worm or virus propagation fairly well~\cite{moore_red}.  Note that in
Fig.~\ref{fig:no_response} the number of malware packets arriving at
the two graphed sub-networks starts small when only one sub-network is initially
infected.  As the worm or virus spreads, the number of inbound
malware packets increases rapidly for a period but eventually levels
off when more and more sub-networks become infected.

The $H^\infty$-optimal controller performs better than every other
controller whenever malware is present, as seen in
Table~\ref{table:cost_ratio_results}. In this case, we choose a
100:1 malware packet to filtering action cost ratio. The resulting
$\gamma^{*}$ is 4.52.

\begin{table}
\renewcommand{\arraystretch}{1.3}
\caption{\label{table:cost_ratio_results} Cost Ratios ($L$) of Controllers
Under Various Attacks ($b = 0.5$)}
\begin{center}
\begin{tabular} {|c|c|c|c|c|c|}
\hline
Attack & R1 & R2 & R3 & R4 & R5 \\
\hline
A1 & 0.00 & 3.48 & 0.00 & 2.35 & 2.04\\
A2 & 8.36 & 3.00 & 8.02 & 4.45 & 5.42\\
A3 & 9.07 & 2.88 & 5.76 & 4.42 & 4.71\\
A4 & 9.42 & 2.90 & 5.31 & 4.49 & 5.15\\
\hline
\end{tabular}
\end{center}
\end{table}

Table~\ref{table:cost_results} shows the actual costs incurred by
the system in each scenario with the same cost structure. The
significantly lower cost values for the $H^\infty$-optimal
controller in the face of attacks highlight its
ability to filter enough to prevent sub-networks from becoming
infected.

\begin{table}
\renewcommand{\arraystretch}{1.3}
\caption{\label{table:cost_results} Costs ($\| z \|^2$) of
Controllers Under Various Attacks ($b = 0.5$) ($\times 10^3$)}
\begin{center}
\begin{tabular} {|c|c|c|c|c|c|}
\hline
Attack & R1 & R2 & R3 & R4 & R5 \\
\hline
A1 & 0     & 1.172 & 0     & 0.788 & 0.682 \\
A2 & 105.4 & 18.24 & 94.08 & 46.85 & 88.24 \\
A3 & 22.68 & 5.579 & 16.77 & 12.50 & 10.34 \\
A4 & 27.97 & 4.979 & 13.51 & 12.63 & 14.24 \\
\hline
\end{tabular}
\end{center}
\end{table}

\begin{figure}
\centering
\includegraphics[width=3.25in]{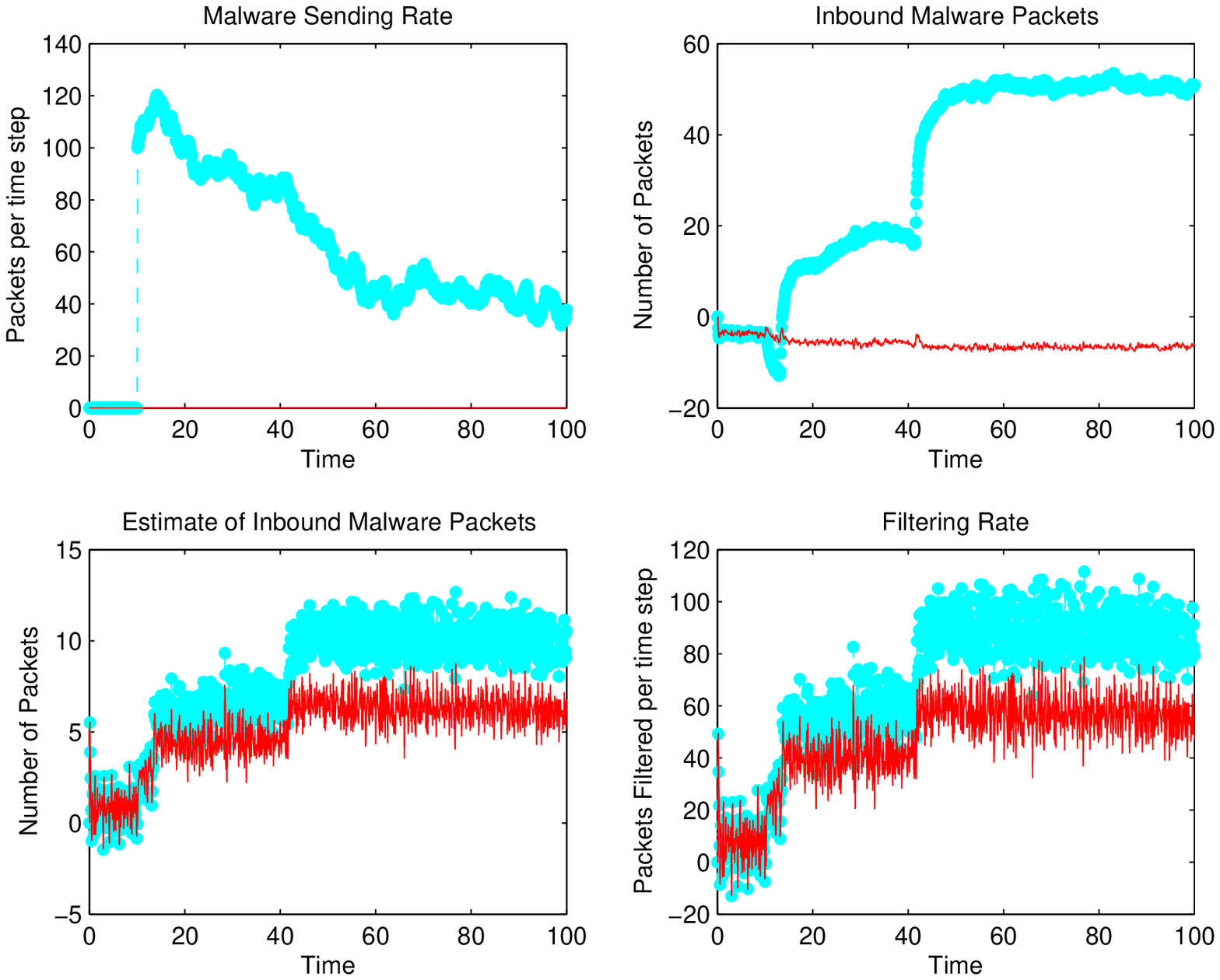}
\caption{\label{fig:resp_hinf_attack_slow} Numerical analysis of
slow worm attack with $H^{\infty}$ response applied on two sub-networks.}
\label{resp_hinf_attack_slow}
\end{figure}

The preventative ability of the $H^{\infty}$-optimal controller can
also be observed in Fig.~\ref{fig:resp_hinf_attack_slow}.  As soon as the
first network detects an increase in inbound malware packets shortly
after 10 time units, the controller begins filtering significantly
(see Fig.~\ref{fig:resp_hinf_attack_slow} ``Filtering Rate") all
across the network.
This prevents the second sub-network
from becoming infected.
We indeed observe that it never sends
malware packets in Fig.~\ref{fig:resp_hinf_attack_slow} ``Malware
Sending Rate."

The ability of the centralized $H^{\infty}$-optimal controller to
respond network-wide to an attack, and hence, increase filtering rates
significantly even on sub-networks where there are not yet many
malware packets being detected, provides an advantage over other
controllers. Another advantage is that it tends to filter
packets aggressively (see Fig.~\ref{fig:resp_hinf_attack_slow}).
We observe this robustness property of the
$H^{\infty}$-optimal controller in the ``Filtering Rate" graph of
Fig.~\ref{fig:resp_hinf_attack_slow}, where the number of packets
filtered is higher than the number of inbound malware packets. This
also contributes to preventing infections, decreasing cost to the
network ($\| \zv \|^2$), and to guaranteeing some level of performance
($\gamma$).

\begin{figure}
\centering
\includegraphics[width=3.25in]{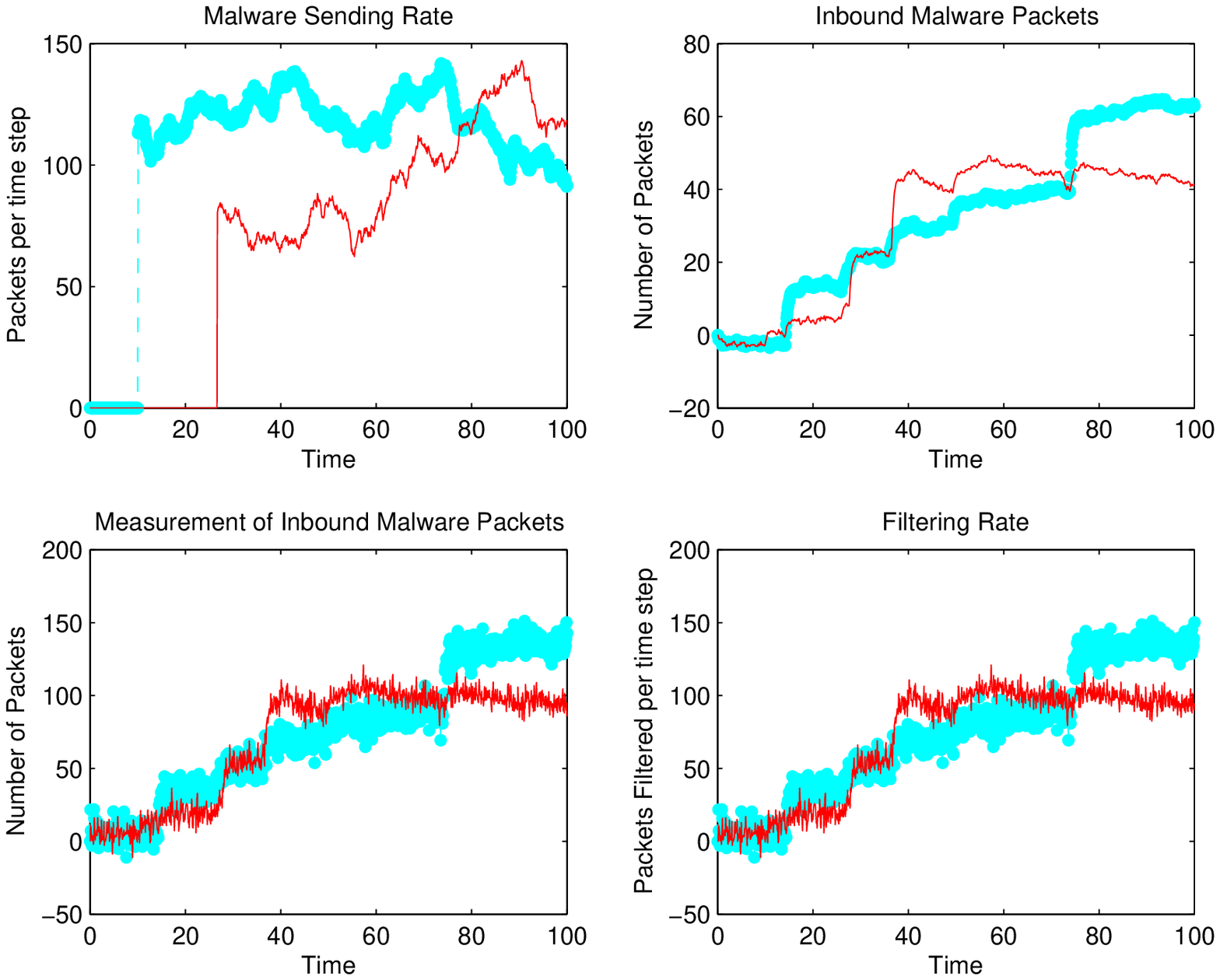}
\caption{\label{fig:resp_rem_det_attack_slow} 
Numerical analysis of slow worm
attack with the controller that removes as many malware
packets as it measures on two sub-networks.} \label{resp_rem_det_attack_slow}
\end{figure}

For comparison, Fig.~\ref{fig:resp_rem_det_attack_slow} shows the
performance of the controller that removes all the estimated malware
packets, thereby disregarding measurement errors and network-wide conditions.  While it does
over-filter, it does \emph{not} filter network-wide when a single
sub-network detects significant numbers of malware packets. Thus, the
uninfected sub-network eventually becomes infected at around time step $25$, which causes it to send malware
(Fig.~\ref{fig:resp_rem_det_attack_slow}).  The LQR optimal controller (R5), on the other hand, does
filter networkwide upon detection of inbound malware packets
anywhere in the network.  It does not, however, filter as much
as the $H^{\infty}$-optimal controller.  Moreover, it is hindered in
that it assumes a zero-mean disturbance, an assumption that becomes
more inaccurate as more sub-networks become infected.

\begin{figure}
\centering
\includegraphics[width=3.25in]{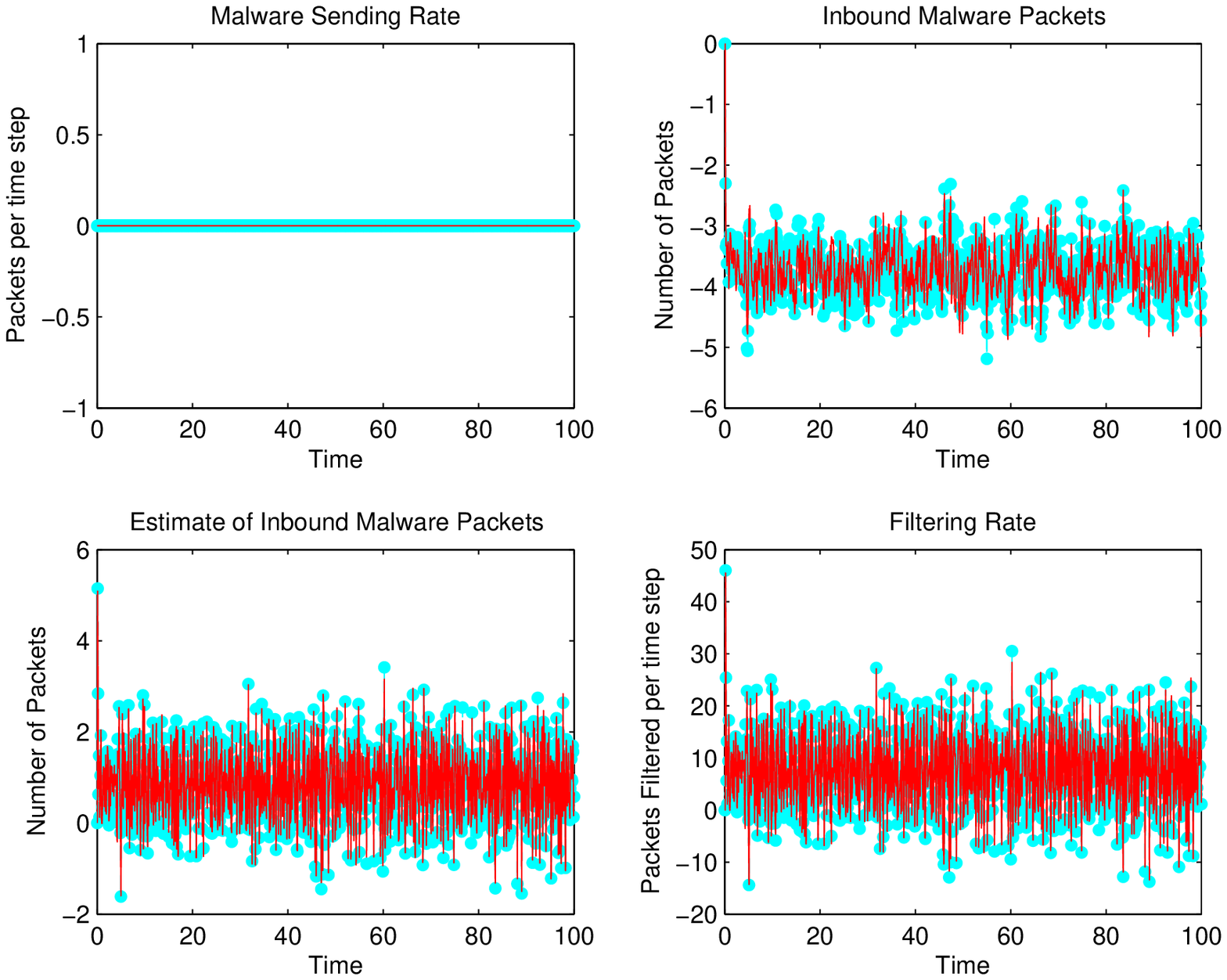}
\caption{\label{fig:resp_hinf_attack_none} Network model response to
no infections with the $H^\infty$-optimal controller.}
\label{resp_hinf_attack_none}
\end{figure}

The $H^\infty$-optimal controller, on the other hand, tends to incur
relatively high costs and cost ratios when there are no infected
sub-networks due to its network-wide over-response (refer to
Tables~\ref{table:cost_ratio_results} and
~\ref{table:cost_results}).  The very
characteristics that make it a strong controller in the face of
attacks prove costly in the absence of attacks.
In fact, the theoretical worst-case
attack is actually quite small in magnitude and essentially
maximizes $L$ by taking advantage of the tiny false alarms and
corresponding excessive filtering that inaccurate measurements
induce in the $H^\infty$-optimal controller.
 Figure~\ref{fig:resp_hinf_attack_none} demonstrates this behavior. Note that the negative number of inbound malware packets indicates that all malware packets have been filtered and legitimate traffic is being removed from the link.

Simulations were also run for other cost functions.  The
$H^\infty$-optimal controller performed relatively better when there
was a greater cost put on the inbound malware packets and vice
versa.  This is to be expected, as this controller is rewarded more
for being cautious when the inbound malware packets increase in cost.  When
the $b$ value was decreased from 0.5 to 0.3, the $H^\infty$-optimal
controller also performs relatively better.  This decrease in $b$
means that when filtering does occur, we are less likely to actually
filter a malicious packet, and thus controllers that filter more are
rewarded.
A decreased $b$ could result from a lower
true-positive rate, a higher false-positive rate, or a higher ratio
of legitimate to malicious traffic.

\subsection{Ns-2 Implementation} \label{sec:ns-2}

We simulate the traffic control algorithm developed at the packet
level using the \textit{Ns-2} network simulator. Our goal is to further
investigate the characteristics of the designed $H^{\infty}$-optimal
controller and demonstrate its capabilities in a realistic setting.
To enable comparisons with the numerical results obtained from
Matlab simulations we define in Ns-2 the same network topology as in
Section~\ref{sec:filtering_numerical}, which is depicted in
Fig.~\ref{fig:system_overview}.
Depending on the specific
application, the end nodes in this graph may represent a sub-network
or any logical or physical set of hosts. As before, we assume high
capacity links between nodes such that no malware packet is dropped
due to congestion, corresponding to a worst-case scenario.
\begin{figure}
\centering
\includegraphics[width=3.25in]{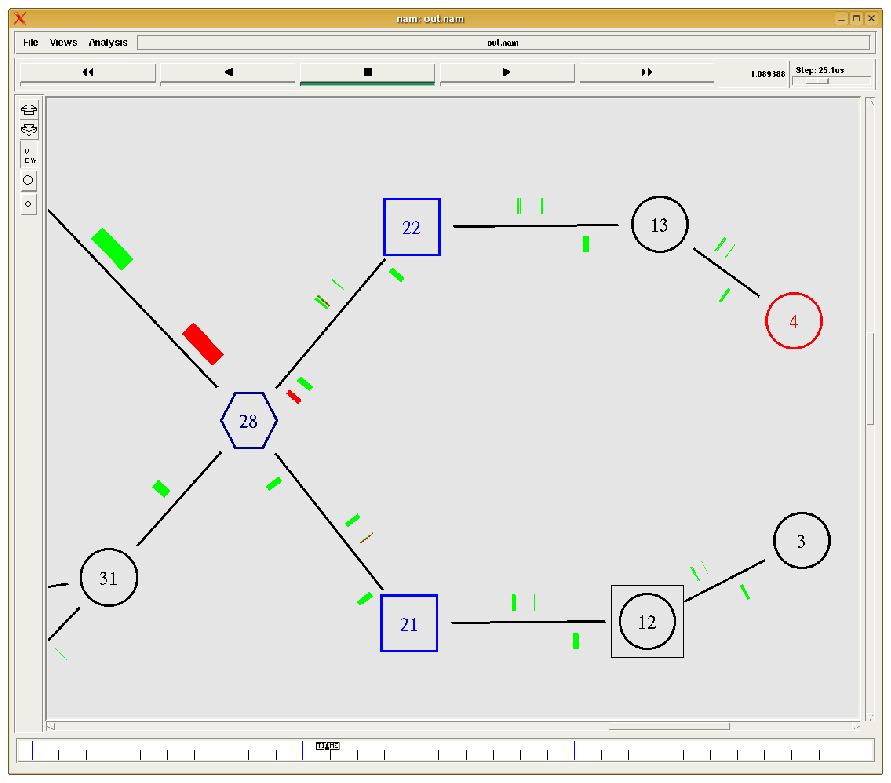}
\caption{\label{fig:ns2screenshot} Screenshot of the Ns-2 simulator
output.  Green packets are legitimate and red packets are malware.}
\label{ns2screenshot}
\end{figure}

In order to simulate the filtering algorithm, we consider here a
specific two-part implementation consisting of \textit{monitoring}
and \textit{filtering} elements. The monitoring nodes, depicted as
hexagons in Fig.~\ref{fig:ns2screenshot}, associate a malware score
$s \in [0,99]$ to each individual packet passing through the link
from the outside. As a simplification, we simulate only inbound
monitoring and filtering. However, a symmetrical outbound
counterpart of the scheme can easily be implemented. The monitoring
elements use this score $s$ and a specific constant threshold to make an
initial estimate on the nature of the packet and label it as malware
or not.  A count of these observed malware packets gives $\yv(t)$.
The monitoring node may utilize
any set of algorithms or approaches to determine this quantity.
We generate the scores randomly according to 
different 
probability distributions for legitimate and malicious packets and use a fixed threshold to simulate this process. This method is similar in some ways to the scoring strategy proposed in ~\cite{kim}.

The filtering elements depicted as boxes in
Fig.~\ref{fig:ns2screenshot} first fetch the malware score $s$ and the
flag from the headers of inbound packets, and then use either a heuristic
or a $H^\infty$ controller-based 
algorithm to make filtering decisions. In this implementation, the
algorithms decide on a time-varying threshold value (different than the previous constant measurement threshold), resulting in a
dynamic filtering scheme. The packets with a score higher than the
threshold are %
filtered. For comparison
purposes, we simulate the R4 algorithm in
Section~\ref{sec:sim_setup}, which we denote as \textit{heuristic}, in addition to
the $H^\infty$ algorithm.
We do not simulate any filtering scheme with a time-invariant threshold as
it clearly would under perform in a dynamic network environment
when compared with the dynamic threshold algorithms.

We calculate the $H^{\infty}$-optimal controller offline in Matlab
and transfer the results to the Ns-2 simulator. In accordance with
the model in Section~\ref{sec:filtering}, the resulting controller
decides on the number of malware packets to be filtered at a given
time interval. %
We translate this number into 
a threshold value by periodically observing the distribution of
scores generated by the monitoring element. Hence, the threshold is chosen such that the number of
packets with a score higher than the threshold (i.e., to be
filtered) matches the number dictated by the
$H^{\infty}$-optimal controller.

\begin{rem} \label{rem:implementation}
It is important to note that the example Ns-2 implementation we
choose here does not play a significant role for the analysis and
demonstration of our algorithm. In fact, depending on the specific
application at hand, one can choose a variety of equivalent
implementations without loss of any generality. For example, the
monitoring and filtering elements can be parts of larger units each
or combined within a dedicated physical device. Or the monitoring
element can be deployed as a dedicated hardware device and the
filtering element as part of a firewall implementation. Clearly, the
possible combinations are numerous.
\end{rem}

We simulate, compare, and contrast the $H^{\infty}$ and
detection-based heuristic filtering schemes in a variety of scenarios under
different cost structures, detection capabilities, and traffic
levels. The hypothetical scenarios we consider are summarized as
follows:
\begin{enumerate}
 \item A cost on malware packets ($\xv$) to cost on filtering ($\uv$) ratio of
 100:1 in $\| \zv \|^2$ and $L$. We assume that the monitoring devices are capable of scoring and labeling only half of the malware packets correctly (S1). 
 \item The cost is the same as in Scenario 1, but we consider a more pessimistic case where the monitoring device only detects a quarter of the total malware
 packets (S2).
 \item This scenario is the same as Scenario 1 except for an increase in the cost coefficient ratio to
 200:1 (S3).
 \item Likewise, this scenario is the same as Scenario 2 with a cost coefficient ratio of 200:1 (S4).
 \item The final scenario matches Scenario 1 but has a cost coefficient ratio of 0.1:1 (S5).
\end{enumerate}

In all of the above cases, each end-node (sub-network) sends
randomly fluctuating $1000$-KB legitimate traffic to all
sub-networks. In addition we consider an ``infection'' or worm-like malware
propagation scheme, where each sub-network becomes ``infected'' with
some probability if it receives sufficiently many malware packets
and afterward generates malware traffic of $200$-KB to other nodes.

\begin{table}[htp]
\renewcommand{\arraystretch}{1.3}
\caption{\label{table:ns2_results} Cost Results of Ns-2 Simulations}
\begin{center}
\begin{tabular} {|c|c|c|c|c|c|}
\hline
 & \multicolumn{3}{c|}{$H^{\infty}$-Optimal} & \multicolumn{2}{c|}{Detection-Based} \\
\hline
Scen. & $L$ & $\z$ \;\;($\times 10^6$) & $\gamma^*$ & $L$ & $\z$ \;\;($\times 10^6$)\\
\hline
S1 & 3.9 & 77 & 3.2 & 4.9 & 147 \\
S2 & 3.7 & 89 & 3.2 & 6.6 & 369 \\
S3 & 4.2 & 87 & 4.2 & 6.9 & 287 \\
S4 & 4.9 & 155 & 4.2 & 9.3 & 736 \\
S5 & 0.31 & 0.68 & 0.3 & 1.05 & 6.78 \\
\hline
\end{tabular}
\end{center}
\end{table}

The numerical results for both of the algorithms under each scenario
described above are summarized in Table~\ref{table:ns2_results}. We
observe here several expected characteristics of the $H^{\infty}$
controller such as optimality with respect to the cost functions and
robustness. In almost all of the cases and over a wide range of cost
coefficient ratios it outperforms the detection-based heuristic scheme.
More importantly, it exhibits robustness with respect to variations
in detection quality (see case 1 versus 2) and guarantees an upper
bound on the cost $L$. It is observed that the $L$ value is always
near the theoretically calculated bound $\gamma^*$.
Another indication of the $H^{\infty}$-optimal controller's
robustness is the satisfactory performance of the controller even
though it is calculated offline with estimated system
characteristics. This, along with the assumptions inherent in the
model, explains the occasional discrepancies observed between $L$ values
and the theoretical upper-bounds $\gamma^*$.

\begin{figure}[htp]
\centering
\begin{tabular}{cc}
 \includegraphics[width=1.5in]{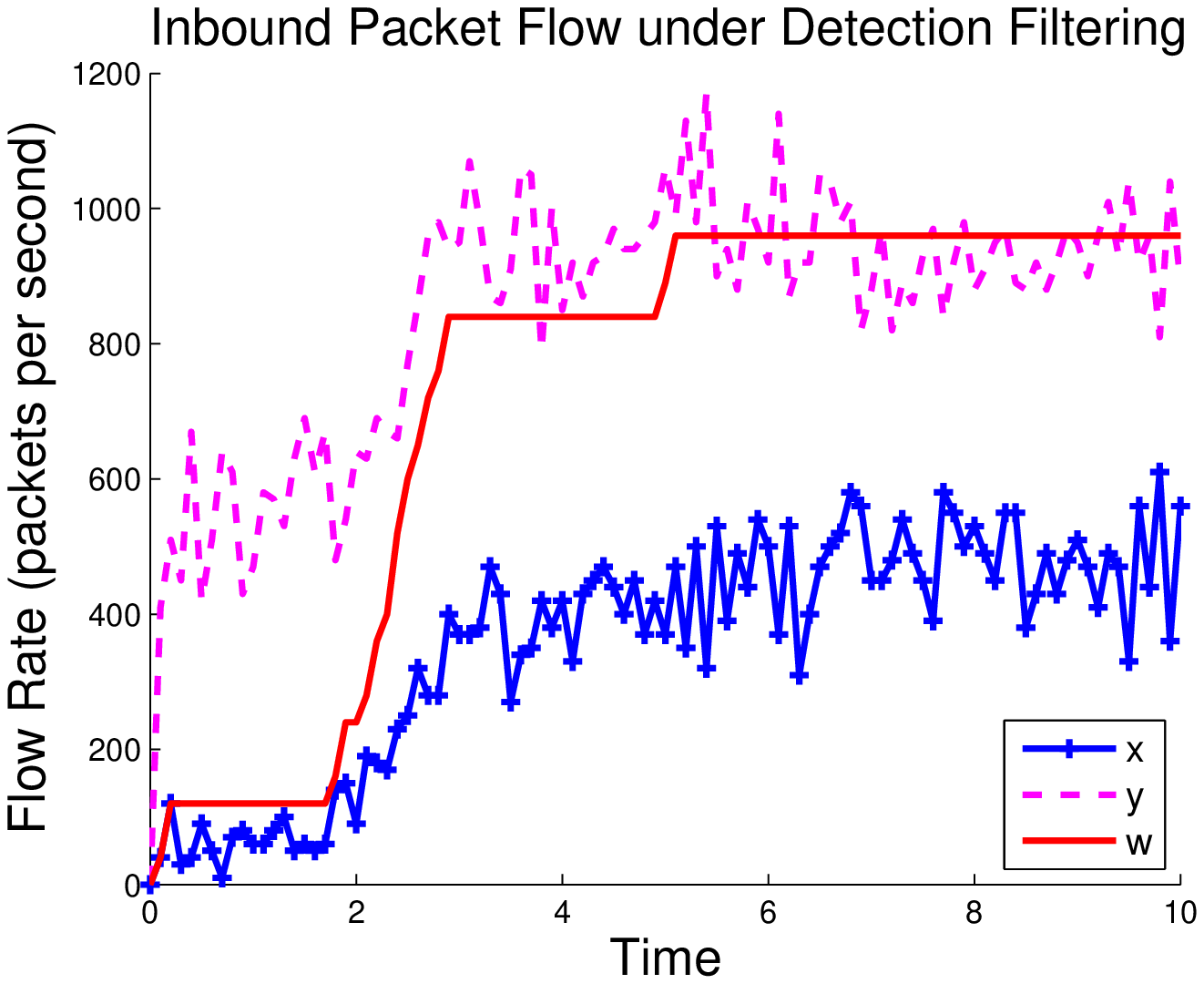}
  & \includegraphics[width=1.5in]{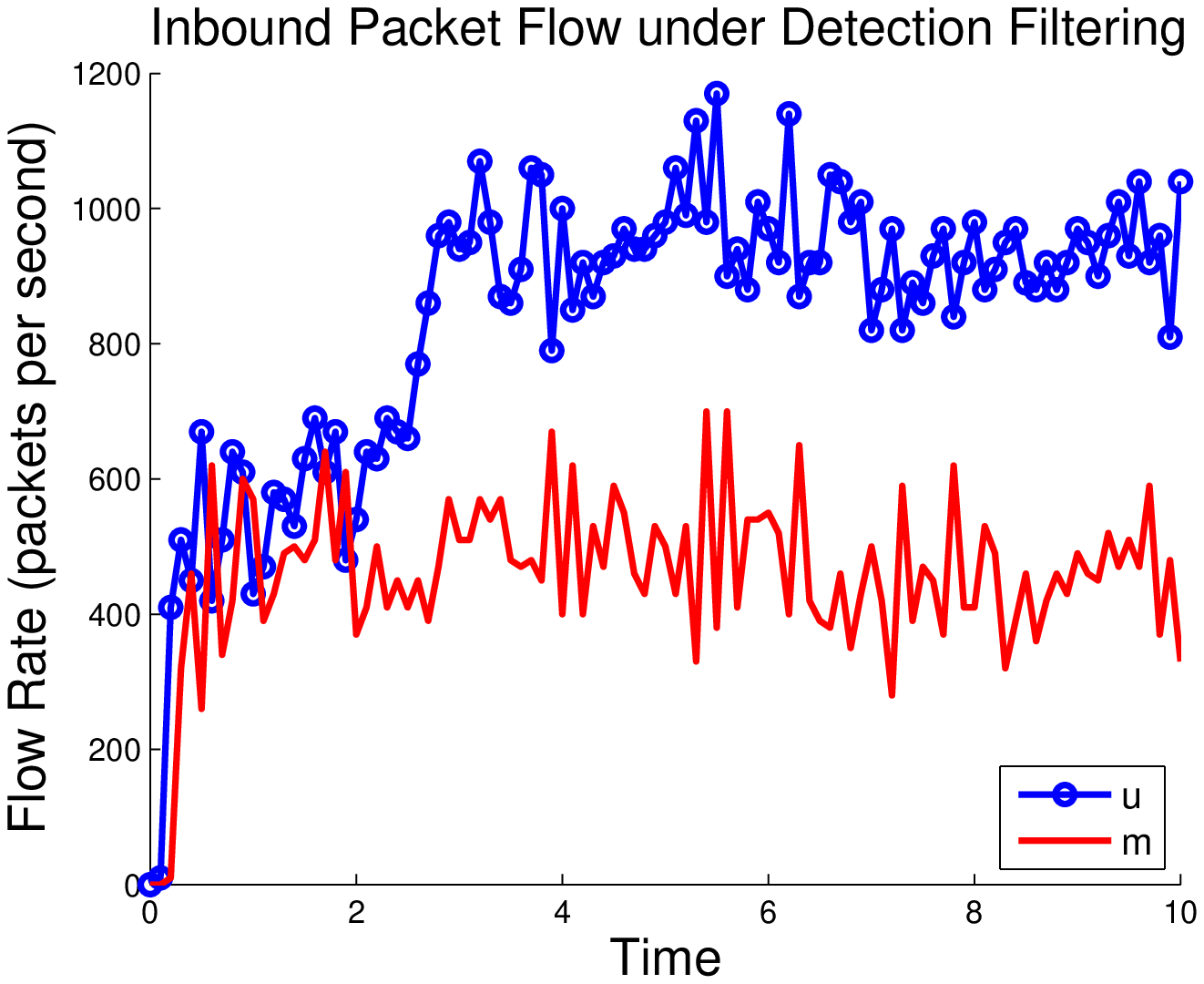}
\end{tabular}
\caption{\label{fig:symtrcost100} Various inbound packet flow rates
to sub-network 1 under the detection-based filtering.}
\end{figure}

\begin{figure}[htp]
\centering
\begin{tabular}{cc}
 \includegraphics[width=1.5in]{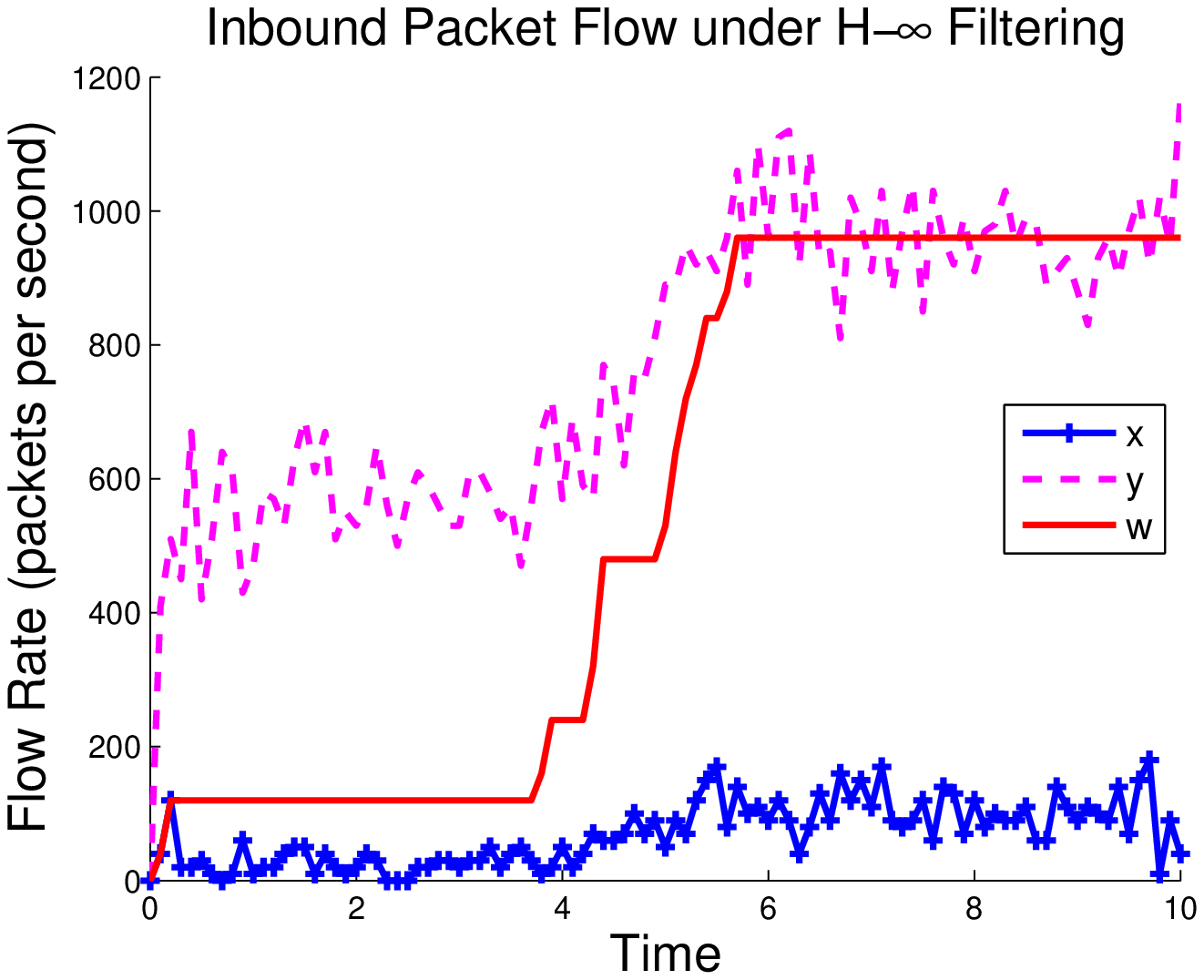}
  & \includegraphics[width=1.5in]{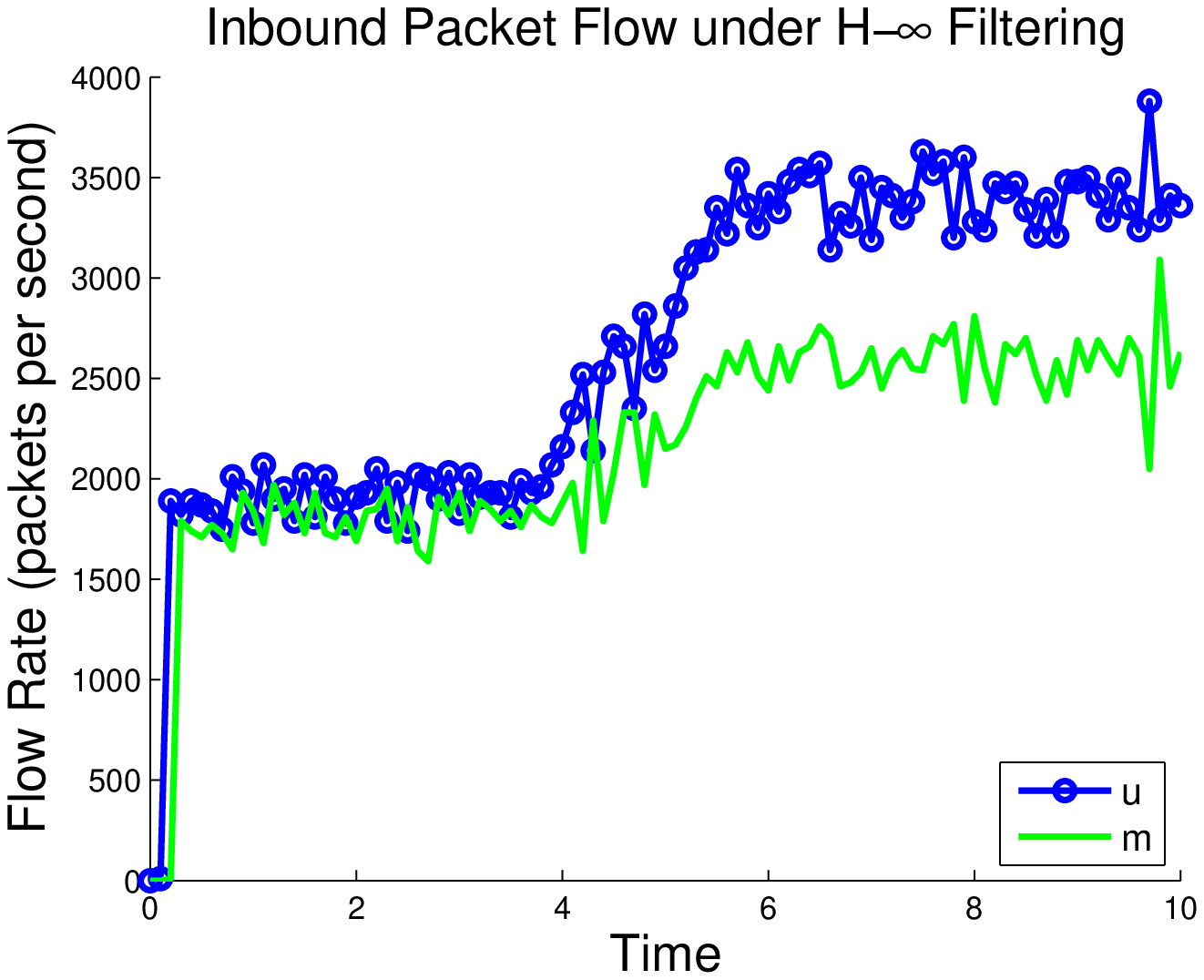}
\end{tabular}
\caption{\label{fig:symhinfcost100} Various inbound packet flow rates to sub-network 1 under  $H^{\infty}$ controller.}
\end{figure}

\begin{figure}[ht]
\centering
\begin{tabular}{cc}
 \includegraphics[width=1.5in]{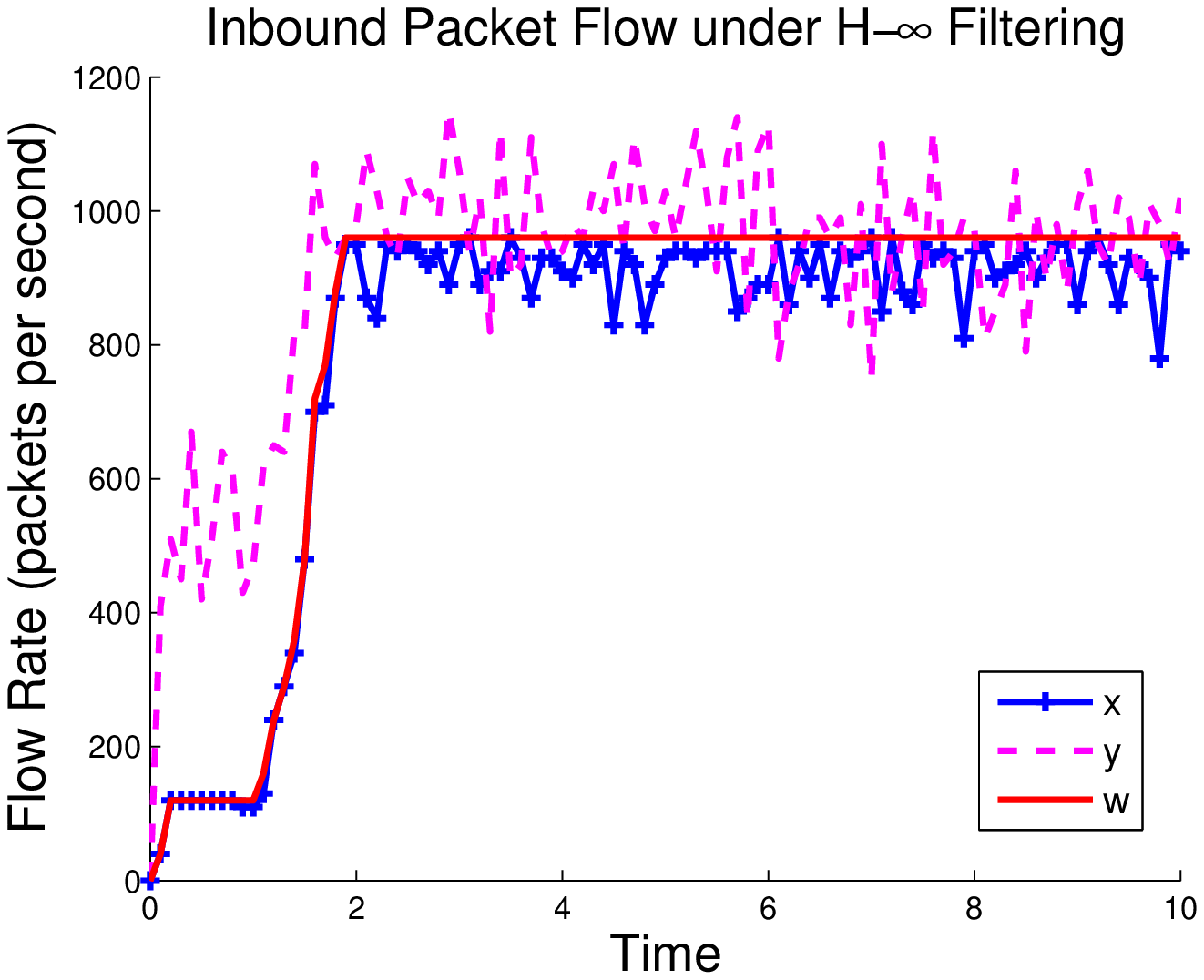}
  & \includegraphics[width=1.5in]{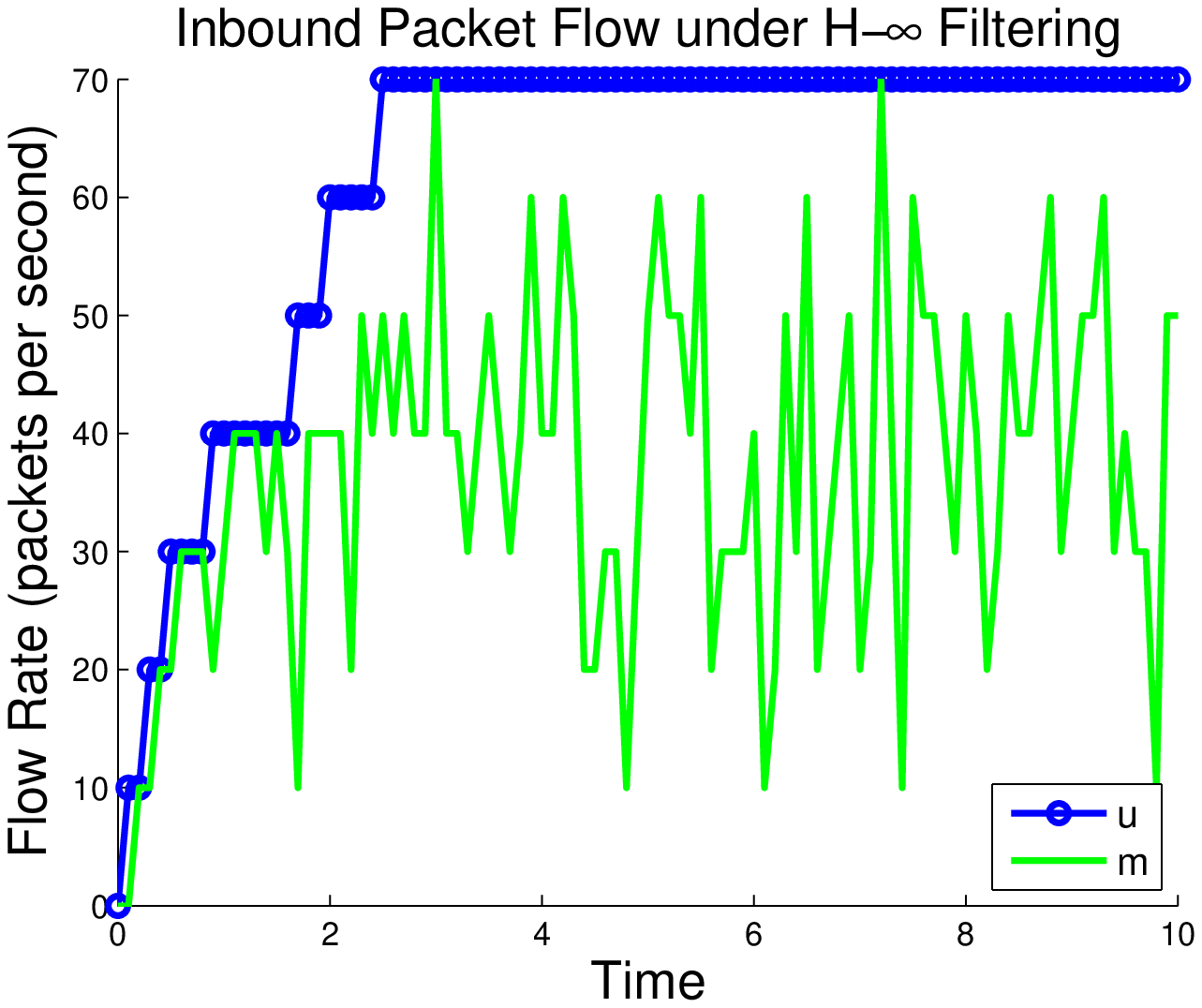}
\end{tabular}
\caption{\label{fig:symhinfcost1} Various inbound packet flow rates to sub-network 1 under  $H^{\infty}$ controller when the cost coefficient ratio is $0.1:1$ (Scenario 5).}
\end{figure}

We next analyze the time-series data collected for a representative
sub-network. We depict various quantities of interest $x$ (malware
packets that pass through the filter), $y$ (packets labeled as
malware by monitor), and $u$ (filtering rate) as in
Sub-section~\ref{sec:matlab}. In addition, we plot the
the rate of falsely positive labeled packets $m$ and the rate of
real malware flow, $w$.
Figure~\ref{fig:symtrcost100}
shows the evolution of these quantities over time in Scenario 1
under the detection-based scheme, whereas
Fig.~\ref{fig:symhinfcost100} depicts the counterpart for the
$H^{\infty}$ controller.
We observe that the
$H^{\infty}$ controller performs better than the detection-based scheme in
terms of removing the malware packets through aggressive filtering
in line with the preferences expressed in the cost function.
Concurrently, this leads to a slower infection rate as can be
inferred from the evolution of real malware flow rate ($w$) in
Fig.~\ref{fig:symhinfcost100}. On the other hand, when the cost
coefficient ratio changes to the one in Scenario 5, $H^{\infty}$
controller is much less aggressive in filtering due to high cost of
dropping legitimate packets.  This can be seen in Fig.~\ref{fig:symhinfcost1}, where the maximum filtering rate is significantly lower that in other scenarios.

\begin{figure}[ht]
\centering
 \includegraphics[width=3.25in]{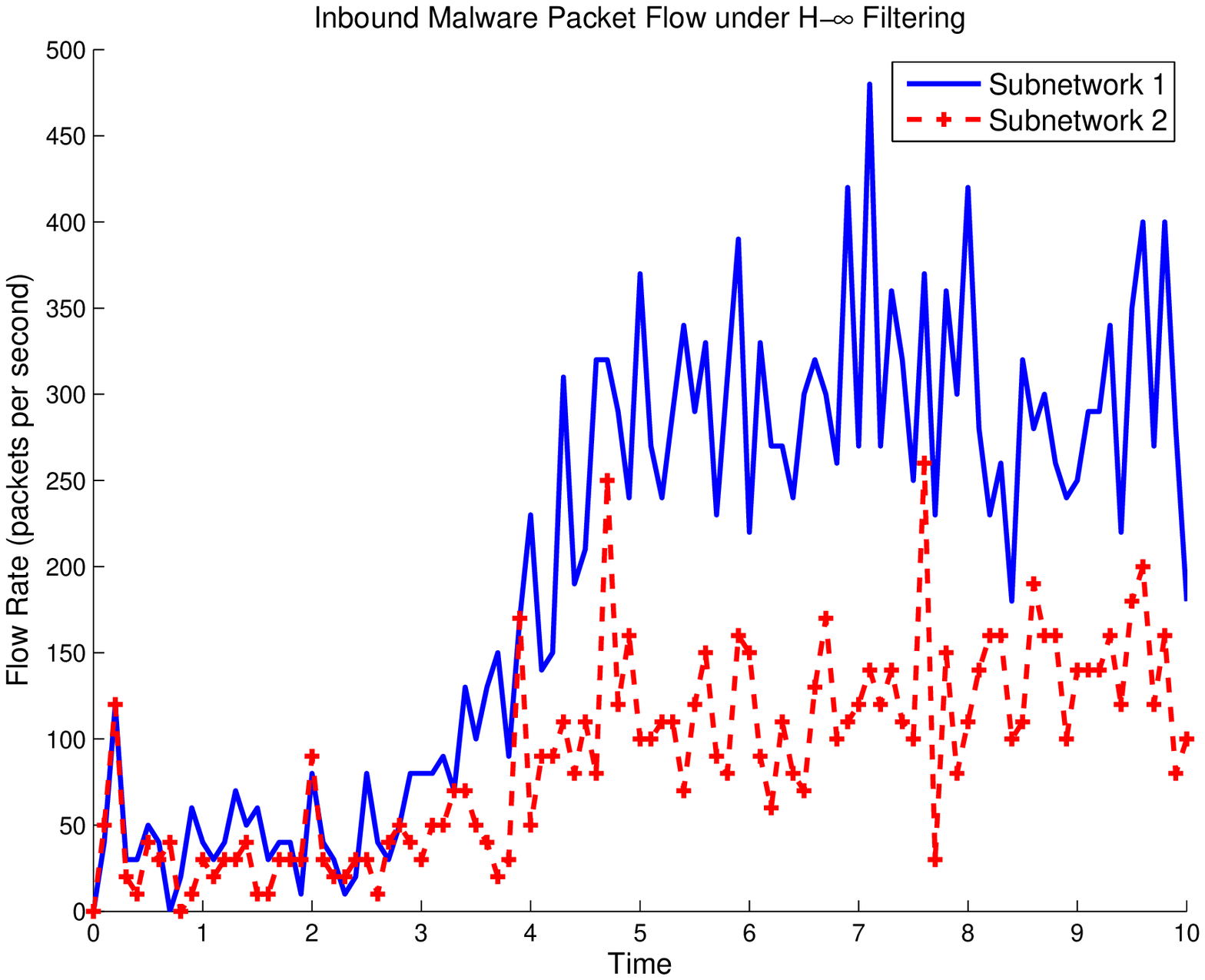}
\caption{\label{fig:asymmetric} Inbound malware packet flow rates to
sub-networks 1  and 5 (more valuable) under  $H^{\infty}$
controller.}
\end{figure}

We finally consider the case when one of the sub-networks (say $5$)
is more valuable than others and needs more intensive inbound
filtering. This preference can easily be reflected to the cost
function by increasing the respective entry of the matrix $H$ in
(\ref{e:hinf_z}). Thus, the $H^{\infty}$ controller reacts
accordingly and filters more aggressively for this sub-network
compared to any other as depicted in Fig.~\ref{fig:asymmetric}.

\section{Conclusion} \label{sec:h_inf_conclusion}

We have studied an application of robust control theory to
network security by investigating an $H^{\infty}$-optimal control formulation of the network filtering problem that captures its inherent challenges such as the base-rate fallacy 
and takes into account relevant costs. The corresponding $H^{\infty}$-optimal controller
has been derived and analyzed numerically in Matlab as well as simulated in Ns-2.  
The controller performs better than alternative controllers when there is a
significant amount of malware traffic present.  In addition, it provides a certain performance
guarantee for a wide range of conditions.

There exist several possible extensions to this work.  Obtaining a
distributed version of this controller for a larger system could be
one future direction. Another research direction is the application
of similar $H^{\infty}$-optimal controllers to other network security problems, such as spam
filtering and DDoS attacks.

\section*{Acknowledgment}
The authors would like to thank the Boeing Corporation and Deutsche Telekom, AG for their support of this research, for the former through the Information Trust Institute at the University of Illinois at Urbana-Champaign. An earlier, more  concise version of this paper will be presented at the 46th IEEE Conference on Decision and Control, New Orleans, December 12-14, 2007, with the title ``An optimal control approach to malware filtering."

\bibliographystyle{IEEEtran}
\bibliography{security}

\begin{thebibliography}{10}
\providecommand{\url}[1]{#1}
\csname url@rmstyle\endcsname
\providecommand{\newblock}{\relax}
\providecommand{\bibinfo}[2]{#2}
\providecommand\BIBentrySTDinterwordspacing{\spaceskip=0pt\relax}
\providecommand\BIBentryALTinterwordstretchfactor{4}
\providecommand\BIBentryALTinterwordspacing{\spaceskip=\fontdimen2\font plus
\BIBentryALTinterwordstretchfactor\fontdimen3\font minus
  \fontdimen4\font\relax}
\providecommand\BIBforeignlanguage[2]{{%
\expandafter\ifx\csname l@#1\endcsname\relax
\typeout{** WARNING: IEEEtran.bst: No hyphenation pattern has been}%
\typeout{** loaded for the language `#1'. Using the pattern for}%
\typeout{** the default language instead.}%
\else
\language=\csname l@#1\endcsname
\fi
#2}}

\bibitem{moore_slammer}
D.~Moore, V.~Paxson, S.~Savage, C.~Shannon, S.~Staniford, and N.~Weaver,
  ``Inside the slammer worm,'' \emph{IEEE Security \& Privacy Magazine},
  vol.~1, pp. 33--39, July-Aug. 2003.

\bibitem{moore_red}
D.~Moore, C.~Shannon, and K.~Claffy, ``Code-red: {A} case study on the spread
  and victims of an internet worm,'' in \emph{Proc. of ACM SIGCOMM Workshop on
  Internet Measurement}, Marseille, France, 2002, pp. 273--284.

\bibitem{axelsson}
S.~Axelsson, ``The base-rate fallacy and its implications for the difficulty of
  intrusion detection,'' in \emph{Proc. of 6th ACM Conference on Computer and
  Communications Security}, Kent Ridge Digital Labs, Singapore, 1999, pp. 1--7.

\bibitem{rohloff_basar}
K.~Rohloff and T.~Ba\c{s}ar, ``The detection of {RCS} worm epidemics,'' in
  \emph{Proc. of ACM Workshop on Rapid Malcode}, Fairfax, VA, 2005, pp. 81--86.

\bibitem{cisco}
\BIBentryALTinterwordspacing
Cisco, ``{NAT} and stateful inspection in {C}isco {IOS} firewall,'' white
  paper, 2006. [Online]. Available:
  \url{http://www.cisco.com/en/US/tech/tk648/tk361/technologies\_white\_paper
  \\ 09186a0080194af8.shtml}
\BIBentrySTDinterwordspacing

\bibitem{basarhinf}
T.~Ba\c{s}ar and P.~Bernhard, \emph{H$^\infty$-Optimal Control and Related
  Minimax Design Problems: A Dynamic Game Approach}, 2nd~ed.\hskip 1em plus
  0.5em minus 0.4em\relax Boston, MA: Birkh\"auser, 1995.

\bibitem{tulloch}
M.~Tulloch, \emph{Microsoft Encyclopedia of Security}.\hskip 1em plus 0.5em
  minus 0.4em\relax Redmond, WA: Microsoft Press, 2003.

\bibitem{hazelhurst}
S.~Hazelhurst, ``A proposal for dynamic access lists for {TCP/IP} packet
  filtering,'' in \emph{Proc. of South African Instutute of Computer Scientists
  and Information Technologists Annual Conference}, Pretoria, South Africa,
  September 2001, http://arxiv.org/abs/cs/0110013.

\bibitem{kim}
Y.~Kim, W.~C. Lau, M.~C. Chuah, , and H.~J. Chao, ``Packetscore: A
  statistics-based packet filtering scheme against distributed
  denial-of-service attacks,'' \emph{IEEE Trans. on Dependable and Secure
  Computing}, vol.~3, no.~2, pp. 141--155, April-June 2006.

\bibitem{zou}
C.~Zou, D.~Towsley, and W.~Gong, ``A firewall network system for worm defense
  in enterprise networks,'' University of Massechusetts, Amherst, MA, Technical
  {R}eport TR-04-CSE-01, Feb. 2004.

\bibitem{chen_quarantine}
T.~M. Chen and N.~Jamil, ``Effectiveness of quarantine in worm epidemics,'' in
  \emph{Proc. of {IEEE} ICC 2006}, Istanbul, Turkey, June 2006, pp. 2142--2147.

\bibitem{alpcancdc04}
T.~Alpcan and T.~Ba\c{s}ar, ``A game theoretic analysis of intrusion detection
  in access control systems,'' in \emph{Proc. 43rd IEEE Conf. Decision and
  Control}, Paradise Island, Bahamas, December 2004, pp. 1568--1573.

\end{thebibliography}

\end{document}